\def\asec{$''$ cy$^{-1}$}
\def\asecc{''\ {\rm cy}^{-1}}
\def\leti{Lense--Thirring}

\def\derp#1#2{\rp{\partial{#1}}{\partial{#2}}}
\def\rfr#1{eq.(\ref{#1})}

\def\leti{Lense--Thirring}
\def\Rfr#1{Eq.(\ref{#1})}

\def\derp#1#2{\rp{\partial{#1}}{\partial{#2}}}

\def\eqi{\begin{equation}}
\def\eqf{\end{equation}}
\def\eqia{\begin{eqnarray}}
\def\eqfa{\end{eqnarray}}

\def\rp#1#2{{#1\over#2}}

\def\ct#1{\cite{#1}}
\def\lb#1{\label{#1}}

\documentclass[11pt]{article}
\usepackage{amsmath,amsthm,amscd,amssymb}
\usepackage{latexsym}
\usepackage{graphics,graphicx}
\usepackage{setspace}
\begin{document}

\noindent{\bf \LARGE{On the possibility of testing  the Dvali-Gabadadze-Porrati brane-world scenario\\
with orbital motions in the Solar System}}
\\
\\
\\
{Lorenzo Iorio}\\
{\it Viale Unit$\grave{a}$ di Italia 68, 70125\\Bari, Italy
\\e-mail: lorenzo.iorio@libero.it}

\begin{abstract}
Recently it has been suggested to use the perigee of the proposed
LARES/WEBER-SAT satellite in order to measure the secular
precession which would be induced on such Keplerian orbital
element by a weak-field modification of gravity occurring in some
Brane-World scenarios put forth by Dvali, Gabadadze and Porrati.
This precession, derived for the first time by Lue and Starkman,
amounts to $\sim\mp 4\times 10^{-3}$ milliarcseconds per year. In
this paper we show that, according to the recently released
EIGEN-CG01C Earth gravity model which combines data from the
dedicated CHAMP and GRACE missions, the quite larger systematic
errors due to the Newtonian part of the terrestrial gravitational
potential would vanish any attempts to detect a so small effect.
Improvements in our knowledge of the Earth's gravitational field
of, e.g., up to four orders of magnitude in the low degree even
zonal harmonics would be required. Also the obtainable
observational accuracy in reconstructing the LARES orbit rules out
the possibility of measuring the LS effect with the perigee of
LARES. The situation is much more favorable in the Solar System
scenario. Indeed, the non-Newtonian perihelion advance of Mars,
which is the currently best tracked planet, has recently been
measured with an accuracy of $1\times 10^{-2}$ milliarcseconds
over a 92-years time span; the LS precession is $\sim\mp 4\times
10^{-1}$ milliarcseconds per century. A suitable combination of
the perihelia and the nodes of Mars and Mercury, which
disentangles the LS effect from the competing larger Newtonian and
general relativistic precessions, might allow to reach a 1-sigma
41$\%$ level of accuracy.
\end{abstract}

\newpage
\tableofcontents
\newpage

\section{Introduction}

The recently observed acceleration of the Universe could be
explained either by invoking a small vacuum energy (or some
effective vacuum energy) or by assuming that the Einstein gravity
undergoes long-range modifications imitating a small cosmological
constant. The second approach seems to be more promising because
it is perturbatively stable under quantum corrections. In it the
conventional gravity breaks down at very large distances, beyond a
certain threshold  $r_{\rm c}$ which is perturbatively stable and
is fixed by the observations to a value $\sim 10^{26}$ m. In such
a framework there is a class of Brane-World theories which, for
$r<<r_{\rm c }$, lead to Einstein-Newton gravity up to small
corrections \ct{Dvali 2004}. As a consequence of such tiny
deviations, a secular advance of the argument of pericentre
$\omega$ of a test particle freely orbiting a central mass $M$
occurs.

According to the Dvali-Gabadadze-Porrati (DVP) model \ct{DVP},
which is a Brane-World theory with an infinite-volume flat extra
spatial dimension, the analytical expression of the pericentre
advance over one orbit of one object moving at distance $r$ from a
central body of mass $M$ is \ct{Lue 2003}
\eqi\Delta\omega=\mp\rp{3\pi\sqrt{2}}{4}\rp{r^{3/2}}{r_{\rm c}
\sqrt{r_{\rm g}}}\ {\rm rad}\ {\rm orbit}^{-1},\eqf where $r_{\rm
c}= 6$ Gpc\footnote{1 parsec (pc)=$3.085678\times 10^{13} $ km.
The value $r_c=6$ Gpc is quoted in \ct{Dvali 2004,
Dva04}}$=1.8514068\times 10^{23}$ km and $r_{\rm g}=2GM/c^2$. The
Lue-Starkman (LS) secular rate for any object orbiting a central
mass along a circular orbit  amounts to $\sim\mp 4\times 10^{-3}$
milliarcseconds per year (mas yr$^{-1}$), or $\mp 4\times 10^{-4}$
arcseconds per century (\asec), only\footnote{Note that in \ct{Lue
2003} $r_c\sim 5$ Gpc is assumed from fitting of both the Cosmic
Microwave Background (CMB) and the type IA supernov${\rm \ae}$
(Starkman, G. 2005; personal communication): this yields $5\times
10^{-4}$ \asec\ for the magnitude of the LS effect. Its sign  is
affected by the cosmological branch \ct{Lue 2003}: it is negative
for the standard cosmological branch, and positive for the
self-accelerated one. }.

The possibility of testing such an effect in the Solar System
scenario has preliminarily been investigated for the first time in
\ct{Lue 2003}. In a recent paper \ct{Ciufolini 2004} it has been
suggested, among other things, to use the perigee of the proposed
Earth artificial satellite LARES in order to detect the LS effect.

In the present work we analyze in more details such possibilities
yielding a realistic evaluation of the error budgets by also
including many relevant sources of systematic errors.

\section{The Earth-LARES system}
The Earth artificial satellite LARES/WEBER-SAT \ct{Ciufolini 1986,
Iorio et al. 2002} was proposed in order to measure the
post-Newtonian general relativistic Lense-Thirring effect on the
orbit of a test particle to a high accuracy level ($\sim 1\%$) in
the gravitational field of the Earth by suitably combining its
data with those of the existing geodetic SLR (Satellite Laser
Ranging) satellites LAGEOS and LAGEOS II. Their orbital parameters
are in Table \ref{para}.
\begin{table}
\caption{Orbital parameters of LAGEOS, LAGEOS II and LARES.
}\label{para}
\begin{tabular}{@{\hspace{0pt}}llll}
\hline\noalign{\smallskip}
 Orbital element & LAGEOS & LAGEOS II & LARES
\\
\noalign{\smallskip}\hline\noalign{\smallskip}
semimajor axis $a$ (km) & $1.2270\times 10^4$ & $1.2163\times 10^4$ & $1.2270\times 10^4$\\
eccentricity $e$ & $4.5\times 10^{-3}$ & $1.4\times 10^{-2}$ & $4.0\times 10^{-2}$ \\
inclination $i$ (deg) & $1.10\times 10^2$ & $5.265\times 10^1$ & $7.0\times 10^1$ \\
\noalign{\smallskip}\hline
\end{tabular}
\end{table}
\subsection{The observational sensitivity} From $\Delta
r=ea\Delta\omega$ \ct{Nordtvedt 2000}, it can be obtained that the
accuracy in measuring the perigee precession over a given
observational time span can be expressed as $\delta\omega=\delta
r/ea$. By assuming a root-mean-square (rms) error of 1 mm in
reconstructing the LARES orbit over, say, one year for a given set
of dynamical force models one gets $\delta\omega=4\times 10^{-1}$
mas.

With\footnote{It may be interesting to note that, with such an
orbital configuration and $i=6.34\times 10^1$ deg the
gravitomagnetic Lense-Thirring precession \ct{leti 1918} would
amount to -1.8 mas yr$^{-1}$ only.} $a=3.6\times 10^4$ km and
$e=2.8\times 10^{-1}$, as also suggested in \ct{Ciufolini 2004},
and by assuming a rms error of 1 cm, the accuracy in the perigee
would amount to $2\times 10^{-1}$ mas. Note that the adopted
values for the obtainable accuracies in $r$ are optimistic; for
example, the mm accuracy has not yet been fully achieved for the
existing LAGEOS satellites.
\subsection{The
systematic errors of gravitational origin} The perigee of an Earth
artificial satellite is affected by various kinds of long-period
(i.e. averaged over one orbital revolution) orbital perturbations
induced by the Newtonian Earth's geopotential \ct{Kaula 1966}. The
most insidious ones are the secular rates induced by the even
($\ell=2,4,6...$) zonal ($m=0$) harmonic coefficients $J_{\ell}$
of the geopotential which account for the departure of the Earth
from an exact spherical shape. Their explicit expressions up to
degree $\ell=20$ can be found, e.g., in \ct{Iorio 2003}. The
largest precession is induced by the Earth's quadrupole mass
moment $J_2$. For a moderate eccentricity its analytical
expression is
\eqi\left.\frac{d\omega}{dt}\right|_{J_{2}}=\rp{3}{2}n\left(\frac{R}{a}\right)^2\rp{J_2}{(1-e^2)^2}
\left(2-\frac{5}{2}\sin^2 i \right),\lb{prece}\eqf where
$n=\sqrt{GM/a^3 }$ is the unperturbed Keplerian mean motion and
$R$ is the Earth's equatorial mean radius.

In \ct{Ciufolini 2004} it is proposed to launch LARES in the so
called frozen-perigee orbit characterized by the critical value of
the inclination, $i=6.34\times 10^1$ deg, for which the
$J_2$-precession of \rfr{prece} vanishes. Moreover, it seems that
Ciufolini suggests to use the LARES data together with those from
LAGEOS and LAGEOS II in order to measure the Lense-Thirring as
well.

We will now show that such proposals are unfeasible.
\subsubsection{The impact of the even zonal harmonics of the
geopotential} Indeed, apart from the fact that the unavoidable
orbital injection errors would prevent to exactly insert LARES in
orbit with the required inclination, it turns out that the impact
of the other uncancelled precessions induced by the even zonal
harmonics  of higher degree, along with their secular variations,
would totally swamp the LS effect. In Table \ref{orpar} we use the
calibrated sigmas of the even zonal harmonics of  the recently
released combined CHAMP+GRACE+terrestrial gravity EIGEN-CG01C
Earth gravity model \ct{Reigber et al. 2004} in order to calculate
the mismodelled residual classical precessions on the perigee of
LARES by assuming $i=6.34\times 10^1$ deg.
\begin{table}
\caption{Mismodelled classical secular precessions, in mas
yr$^{-1}$, of the perigee of an Earth satellite with semimajor
axis $a=1.2270\times 10^4$ km, inclination $i=6.34\times 10^1$
deg, eccentricity $e=4.0\times 10^{-2}$ according to the combined
CHAMP+GRACE+terrestrial gravity EIGEN-CG01C Earth gravity model up
to degree $\ell=20$ (1-sigma). For the precessions induced by the
secular variations of the even zonal harmonics, referred to an
observational time span of one year, the values for $\dot
J_{\ell}, \ell=2,4,6$ of \ct{Cox et al. 2003} have been used. Such
precessions grow linearly in time.  In the last three rows a - has
been inserted because the corresponding mismodelled precessions
amount to $10^{-4}$ mas yr$^{-1}$. The LS secular precession
amounts to $\mp 4\times 10^{-3}$ mas yr$^{-1}$ for an object
orbiting a central mass along a circular path. }\label{orpar}

\vspace{0.5cm}
\begin{tabular}{@{\hspace{0pt}}ll}
\hline\noalign{\smallskip} Even zonal harmonic & Mismodelled
precessions (mas yr$^{-1}$)
\\
\noalign{\smallskip}\hline\noalign{\smallskip}
$J_2$ & $1.26\times 10^{-1}$\\
$\dot J_2$ & $4\times 10^{-3}$ mas yr$^{-2}$\\
$J_4$ & 4.390\\
$\dot J_4$ & 1.406 mas yr$^{-2}$\\
$J_6$ & 1.355\\
$\dot J_6$ & $6.63\times 10^{-1}$ mas yr$^{-2}$\\
$J_8$ & $2.08\times 10^{-1}$\\
$J_{10}$ & $2.5\times 10^{-2}$\\
$J_{12}$ & $2.5\times 10^{-2}$\\
$J_{14}$ & $7\times 10^{-3}$\\
$J_{16}$ & -\\
$J_{18}$ & -\\
$J_{20}$ & -\\

\noalign{\smallskip}\hline
\end{tabular}
\end{table}
It can be easily seen that the mismodelled precessions induced by
the first seven even zonal harmonics are larger than the LS
precession. Due to the extreme smallness of such an effect, it is
really unlikely that the forthcoming Earth gravity models from
CHAMP and GRACE will substantially change the situation. This
rules out the possibility of using only the perigee of LARES.

\subsubsection{The linear combination approach} In regard to the
possibility of suitably combining the Keplerian orbital elements
of the existing LAGEOS satellites and of the proposed LARES
\ct{Ciufolini 1996, Iorio 2004, Iorio and Doornbos 2004} in order
to reduce the impact of the systematic errors of the geopotential
on the proposed measurement, it is unfeasible as well.

Indeed, the perigee is also affected, among other things, by the
post-Newtonian general relativistic Einstein precession
\ct{Einstein 1915}, whose nominal value for LARES is
$3.280136\times 10^3$ mas yr$^{-1}$, and by the Lense-Thirring
effect, which, for $i=6.34\times 10^1$ deg, nominally amounts to
-$4.1466\times 10^1$ mas yr$^{-1}$. This means that if we want to
measure the LS precession independently of such quite larger
Newtonian and post-Newtonian effects we would need ten Keplerian
orbital elements in order to write down a linear system of ten
equations in ten unknowns (the first seven even zonal harmonics,
the LS effect and the two relativistic precessions) and solve it
for the LS precession. Instead, we would have at our disposal, in
principle, only four reliable Keplerian orbital
elements\footnote{Note that the other routinely and accurately
laser-tracked  SLR satellites which could, in principle, be
considered are Ajsai, Starlette and Stella: the useful orbital
elements are their nodes and the perigee of Starlette. However,
since they orbit at much lower altitudes than the LAGEOS
satellites they would practically be useless. Indeed, they are
sensitive to much more even zonal harmonics of the geopotential so
that they would greatly increase the systematic error induced by
them.}: the longitudes of the ascending node $\Omega$ of LAGEOS,
LAGEOS II and LARES-which are affected by the Earth's geopotential
and by the Lense-Thirring effect but not by the LS force-and the
perigee of LARES. They would only allow to cancel out the general
relativistic effects and just one even zonal harmonic.

The data from an hypothetical satellite with $a=3.6\times 10^4$
km, as proposed in \ct{Ciufolini 2004}, could not be used for the
following reasons. If, on the one hand, the classical geopotential
precessions would be smaller than the LS effect, apart from those
induced by $J_4$ and $J_6$ according to EIGEN-CG01C, on the other
hand, the gravitomagnetic Lense-Thirring and the gravitoelectric
Einstein post-Newtonian precessions on the perigee would amount to
-1.8512 mas yr$^{-1}$ (for $i=6.34\times 10^1$ deg) and
$2.409956\times 10^2$ mas yr$^{-1}$, respectively. This means
that, even for such a higher altitude, one could not analyze only
the perigee of LARES whose Keplerian orbital elements should be,
instead, combined with those of the existing LAGEOS and LAGEOS II:
with the nodes of the LARES, LAGEOS and LAGEOS II and the perigee
of LARES it would be possible, in principle, to disentangle the LS
effect from the post-Newtonian precessions and one classical even
zonal harmonic. For such a high altitude the period of the node of
LARES would amount to $\sim 10^4$ days, i.e. tens of years. The
tesseral $K_1$ tidal perturbation, which is one of the most
powerful harmonic time-dependent perturbations which are not
cancelled out by the linear combination approach, has just the
period of the node. Then, it would act as a superimposed linear
bias over an observational time span of a few years. Moreover, as
shown in \ct{Iorio 2002, Vespe and Rutigliano 2004}, when
high-altitude satellites are included in linear combinations
involving also lower satellites as the existing LAGEOS and LAGEOS
II it turns out that the orbital elements of the higher SLR
targets enter the combination with huge coefficients which amplify
all the uncancelled orbital perturbations. This would also be the
case for the $K_1$ tide affecting LARES.
\subsubsection{The impact of the odd zonal
harmonics} The perigee of an Earth artificial satellite is also
affected by long-period harmonic perturbations induced by the odd
($\ell=3,5,7...$) zonal ($m=0$) harmonics of the geopotential. The
largest perturbation is induced by $J_3$: it has a sinusoidal
signature with the period of the perigee. Its analytic expression
is \ct{J3}
\begin{eqnarray}
\left.\rp{d\omega}{dt}\right|_{J_{3}}&=&-\rp{3}{2}n\left(\frac{R}{a}\right)^3\rp{J_3\sin\omega}{e\sin
i (1-e^2)^3}\left[\left(\rp{5}{4}\sin^2 i-1\right)\sin^2
i+\right.\nonumber\\
&+&\left.e^2\left(1-\rp{35}{4}\sin^2 i\cos^2
i\right)\right].\lb{j3}
\end{eqnarray}
 For $i=6.34\times 10^1$ deg the period
of the perigee of LARES, given by \rfr{prece}, is of the order of
$10^5$ days; moreover, the second term of the right-hand-side of
\rfr{j3} does not vanish. This means that the perigee of LARES in
the critical inclination would be affected by an additional
semisecular bias due to $J_3$ which, over an observational time
span of some years, would resemble a superimposed linear trend.
According to EIGEN-CG01C, its mismodelled effect would be $\leq
5\times 10^{-1}$ mas yr$^{-1}$.

This additional bias should be accounted for both in the
perigee-only scenario and in the linear combinations scenario.
Note also that for $a=3.6\times 10^4$ km and $i=6.34\times 10^1$
deg the period of the perigee would amount to $\sim 10^7$ days,
i.e. $\sim 10^4$ years.
\section{The Earth-Moon system}
In \ct{Dvali 2004} the possibility of analyzing the perigee of the
Moon ($a=3.84\times 10^5$ km, $e=0.05490$) is considered in view
of the possible future improvements in the range residuals
accuracy from the cm to the mm level in the the Lunar Laser
Ranging (LLR) technique \ct{LLR 2004}. By assuming $\delta r\sim
1$ mm, the uncertainty in determining the Moon perigee would
amount to $\delta\omega=9\times 10^{-3}$ mas which is the
magnitude of the LS shift over almost 2 years. The currently
available accuracy in LLR range residuals is $\sim 2$ cm which
maps into a $1.8\times 10^{-1}$ mas accuracy on the perigee.
\section{The Solar System scenario}
If we look at the planetary motions in the gravitational field of
the Sun the situation is more favorable. The LS perihelion secular
rates for all the planets, by neglecting the eccentricity of their
orbits, amounts to\footnote{Departures from such value due to the
eccentricities are very likely too small to be detectable, as can
be inferred from the subsequent discussion.} $\mp 4\times 10^{-4}$
\asec. It is undoubtedly a small value if compared, e.g, to the
well known Einstein perihelion advance which, for Mercury, amounts
to $4.29811\times 10^1$ \asec.
\subsection{The observational sensitivity}
Contrary to the Earth-LARES scenario, the braneworld effect lies
at the edge of the present-day available orbit determination
accuracy.
\begin{table}
  \caption{Orbital parameters of Mercury, Venus and Mars
(http://nssdc.gsfc.nasa.gov/planetary/factsheet/). For the
Astronomical Unit (A.U.) we use the value 1 A.U.=149597870691 m.
The angle $\epsilon$ refers to the inclination to the ecliptic:
the inclination $i$ to the reference plane is almost
$\epsilon/2$.}
     \label{planparam}
\vspace{0.5cm}
\begin{tabular}{@{\hspace{0pt}}llll}
\hline\noalign{\smallskip}
Planet  & $a$ (A.U.) & $\epsilon$ (deg) & $e$ \\
\noalign{\smallskip}\hline\noalign{\smallskip}
Mercury & $3.8709893\times 10^{-1}$ &  7.00487 & $2.0563069\times 10^{-1}$ \\
Venus & $7.2333199\times 10^{-1}$ & 3.39471 & $6.77323\times 10^{-3}$ \\
Mars & 1.52366231 & 1.85061 & $9.341233\times 10^{-2}$ \\
\noalign{\smallskip}\hline
\end{tabular}
\end{table}

The best tracked planet is Mars thanks to the various spacecraft
missions to it: Viking (1976-1982), Pathfinder landers (1997),
Mars Global Surveyor (MGS) (1998-2003) and Odyssey (2002-2003).
According to \ct{pitast}, the rms residuals of ranging for Viking
are 8 m, for Pathfinder are 4.4 m, for MGS and Odyssey are 1.4 m.
%
From the results of Table 4 of \ct{pitast} for the nonsingular
elements $h$ and $k$, based on the processing of about 300000
position observations (1911-2003) of different types used for the
EPM2004 ephemerides, it turns out that the formal standard
deviation in the Martian perihelion advance is $1.5\times 10^{-2}$
mas; the braneworld shift over a 92-years time span amounts to
$3.68\times 10^{-1}$ mas.

The present-day situation for Mercury and, especially, Venus is
less favorable mainly due to the systematic errors induced by the
mismodelling in the planetary topography and the small value of
the eccentricity of the venusian orbit. Indeed, from \ct{pitast}
the rms residuals of ranging from all radiometric data (1911-2003)
amount to 1.4 km and 0.7 km for Mercury and Venus, respectively.
The formal standard deviations in the perihelia advance are
$7.6\times 10^{-1}$ mas and $8.772$ mas, respectively.
%
However, the future Messenger (launched in August 2004 by NASA)
BepiColombo and Venus Express (ESA) missions should improve, among
other things , also the accuracy of the ephemerides of Mercury and
Venus.


\subsection{The systematic errors}\lb{syserr}
However, as in the case of the Earth-LARES scenario, the
systematic errors of gravitational origin represent a very
limiting factor. In Table \ref{solsys} the nominal values of the
major Newtonian and post-Newtonian general relativistic competing
precessions are listed.
\begin{table}
\caption{Nominal values, in \asec, of the Newtonian and
post-Newtonian general relativistic secular precessions affecting
the perihelia and the nodes of the inner planets of the Solar
System. For a given planet, the precession labelled with N-body is
due to the classical interactions with the other bodies of the
Solar System, that labelled with GE is due to the post-Newtonian
gravitoelectric Schwarzschild component of the solar gravitational
field, that labelled with $J_2$ is due to the classical effect of
the Sun's quadrupole mass moment $J_2$ and that labelled with LT
is due to the post-Newtonian gravitomagnetic Lense-Thirring
component of the solar gravitational field. The LS perihelion
secular rates for all the planets, by neglecting the
eccentricities of the orbits, amount to $3.95\times 10^{-4}$
\asec. For the numerical calculations of the planetary N-body
precessions see
http://ssd.jpl.nasa.gov/elem$\_$planets.html$\#$rates. For the
Sun's proper angular momentum $J$, which is the source of the
gravitomagnetic field, the value $1.9\times 10^{41}$ kg m$^2$
s$^{-1}$ \ct{Pijpers 2003} has been adopted. For the Sun-planetary
data see \ct{Seidelmann 1992}. } \label{solsys}
\vspace{0.5cm}
\begin{tabular}{@{\hspace{0pt}}llll}
\hline\noalign{\smallskip} Secular rate & Mercury & Venus & Mars
\\
\noalign{\smallskip}\hline\noalign{\smallskip}
$\dot\omega_{\rm N-body}$ & $1.019036\times 10^3$ & $8.87652\times 10^2$ & $2.580836\times 10^3$\\
$\dot\Omega_{\rm N-body}$ & $-4.4630\times 10^2$ & $-9.9689\times 10^2$ & $-1.02019\times 10^3$\\
$\dot\omega_{\rm GE}$ & $4.2981\times 10^1$ & $8.624$ & $1.351$ \\
$\dot\Omega_{\rm GE}$ & $0$ & $0$ & $0$ \\
$\dot\omega_{J_{2}}$ & $5.0609\times 10^{-2}$ & $5.223\times 10^{-3}$ & $3.92\times 10^{-4}$\\
$\dot\Omega_{J_{2}}$ & $-2.5375\times 10^{-2}$ & $-2.613\times 10^{-3}$ & $-1.96\times 10^{-4}$\\
$\dot\omega_{\rm LT}$ & $-3.018\times 10^{-3}$ & $-4.31\times 10^{-4}$ & $-4.5\times 10^{-5}$ \\
$\dot\Omega_{\rm LT}$ & $1.008\times 10^{-3}$ & $1.44\times 10^{-4}$ & $1.5\times 10^{-5}$ \\
 \noalign{\smallskip}\hline
\end{tabular}
\end{table}
It turns out that the accuracy with which the various competing
effects\footnote{Note that the nominal values of the
Lense-Thirring precessions are one order of magnitude smaller than
the LS effect, so that the gravitomagnetic force does not pose
problems.}. are presently known  would make difficult to use only
one perihelion, apart from, perhaps, Mars. The major limiting
effects are the classical N-body precession, the post-Newtonian
general relativistic gravitoelectric
perihelion advance and the classical secular precession due to the
Sun's even zonal harmonic $J_2$.

A very limiting systematic bias might be, in principle, that
induced by the huge Newtonian N-body secular precessions whose
nominal values are up to seven orders of magnitude larger than the
effect of interest. In regard to their mismodelling, the major
source of uncertainty is represented by the $GM$ of the perturbing
bodies among which Jupiter plays a dominant role, especially for
Mars. According to \ct{Jacobson 2003}, the Jovian $GM$ is known
with a relative accuracy of $10^{-8}$; this would imply for the
red planet a mismodelled precession induced by Jupiter of the
order of $10^{-5}$ \asec.
The $GM$ of Saturn is known with a $1\times 10^{-6}$ relative
accuracy \ct{Jacobson 2004}. However, the ratio of the precession
induced on the Mars perihelion by Saturn to that induced on the
Mars perihelion by Jupiter is proportional to $(M_{\rm Sat}/M_{\rm
Jup})(a_{\rm Jup}/a_{\rm Sat })^3\sim 5\times 10^{-2}$. This would
assure that also the effect of Saturn is of the order of $10^{-5}$
\asec. The situation with the precessions induced by Uranus and
Neptune on the martian perihelion is even more favorable. Indeed,
for Uranus $(M_{\rm Ura}/M_{\rm Jup})(a_{\rm Jup}/a_{\rm Ura
})^3\sim 1\times 10^{-3}$ and the relative uncertainty in the
uranian $GM$ is $2\times 10^{-6}$ \ct{Jacobson 1992}. For Neptune
$(M_{\rm Nep}/M_{\rm Jup})(a_{\rm Jup}/a_{\rm Nep })^\sim 3\times
10^{-4}$ and $\delta(GM)/GM=2\times 10^{-6}$ \ct{Jacobson 1991}.
In regard to the precessions induced by the most important
asteroids, their masses are not known with great accuracy
\ct{Standish 2002}; according to recent estimations, their
mismodelled effect would amount to $\sim 10^{-4}$ \asec for the
secular terms (see Appendix \ref{asteroids}).

In regard to the post-Newtonian relativistic perihelion advance,
it is currently known at a $10^{-4}$ level of relative accuracy
from the PPN parameters $\beta$ and $\gamma$ \ct{pitast}.
\begin{table}

\caption{First row: Formal standard deviations, in \asec, in the
measurement of the post-Newtonian secular rates of  the planetary
perihelia from the processing of the ranging observations (E.V.
Pitjeva, private communication, 2004). They have been obtained by
including also the perturbing effects of the asteroids  which are
particularly important for Mars (see Appendix \ref{asteroids}).
The uncertainty in the perihelion advances is one-to-three orders
of magnitude larger than the LS effect for Mercury and Venus,
respectively, but it is slightly smaller than the LS precession
for Mars. Realistic errors may be an order of magnitude larger.
Second row:  Formal standard deviations, in \asec, in the
numerical propagation of the planetary nodal rates averaged over
200 years from the DE405 ephemerides (E.M. Standish, private
communication, 2004). } \label{sensi}
\vspace{0.5cm}
\begin{tabular}{@{\hspace{0pt}}llll}
\hline\noalign{\smallskip}
Secular rate & Mercury & Venus & Mars
\\
\noalign{\smallskip}\hline\noalign{\smallskip}
$\dot\omega$ & $8.6\times 10^{-3}$ & $1.037\times 10^{-1}$ & $1\times 10^{-4}$\\
$\dot\Omega$ & $1.82\times 10^{-4}$ & $6\times 10^{-6}$ & $1\times 10^{-6}$\\
\noalign{\smallskip}\hline
\end{tabular}
\end{table}
It turns out that the accuracy with which the post-Newtonian
general relativistic gravitoelectric perihelion advances can be
determined is one-to-three orders of magnitude larger than the LS
effect for Mercury and Venus, respectively, but it is slightly
smaller than the LS precession for Mars. The figures of Table
\ref{sensi} for the perihelia have been obtained from the
processing of about 300000 position observations (1911-2003) of
different types: also the perturbing effects of the asteroids
\ct{pitast}, which are particularly important for Mars (see
Appendix \ref{asteroids}), have been accounted for. However, it
must be noted that the realistic errors in such figures may be one
order of magnitude larger, so that the mismodelling in the
relativistic effect would tend to mask the LS rate.

The mismodelled precessions induced by $J_2$, by assuming a
relative uncertainty of 10$\%$  in it \ct{pitast}, would be larger
than the LS effect for Mercury and Venus; for Mars such competing
classical feature would, instead, be smaller than the LS effect by
one order of magnitude.

These evaluations would suggest that in  a more or less near
future it might, perhaps, be possible to analyze the perihelion of
Mars only. However, note that a sort of a priori `imprint' of the
LS signature in the recovery of all the other effects affecting
the perihelion of Mars would be present: it might make unreliable
the proposed measurement by driving the outcome of the test just
towards the desired result. Then, a better approach would consist
in disentangling the LS effect from all the other competing
effects. This can be done by suitably combining the perihelion of
Mars with other Keplerian orbital elements affected by such
Newtonian and post-Newtonian precessions. A similar approach has
recently been adopted, in a Solar System context, in \ct{Iorio
PPN, SolSysLT}. By writing down the rates of the perihelia and the
nodes of Mars and Mercury in terms of the LS, general
relativistic, $J_2$ and N-Body precessions it is possible to
consider them as a linear nonhomogeneous algebraic system of four
equations in four unknowns and to solve it with respect to the LS
effect. The so obtained linear combination of the residuals of the
rates of the perihelia and the nodes of Mercury and Mars is, by
construction, independent of the effects induced by the Newtonian
N-body and solar precessions and by the general relativistic
gravitoelectric force. The result is \eqi\delta\dot\omega^{\rm
Mars }+c_1\delta\dot\Omega^{\rm Mars}+c_2\delta\dot\omega^{\rm
Mercury}+c_3\delta\dot\Omega^{\rm Mercury}=X_{\rm
Dv},\lb{combo}\eqf where \eqi c_1=2.5275,\ c_2=-3.14\times
10^{-2},\ c_3=-6.67\times 10^{-2}, \lb{coeffi}\eqf and \eqi X_{\rm
Dv}=\mp 3.87\times 10^{-4}\ \asecc.\eqf It is important to note
that the impact of Mercury is relatively small, as can be inferred
from \rfr{coeffi}. Note that \rfr{combo} does not cancel out the
Lense-Thirring effect: however, according to Table \ref{solsys},
its nominal impact amounts to $2\times 10^{-5}$ \asec. According
to the results for the nonsingular elements $h,\ k,\ p,\ q$ of
Table 4 of \ct{pitast}, The formal 1-sigma error affecting
\rfr{combo} can be evaluated as $41\%$.

\section{Conclusions}
In this paper we have investigated the possibility of testing the
multidimensional brane-world scenario proposed by Dvali, Gabadadze
and Porrati by analyzing in the weak-field regime of the Solar
System the pericentre advance derived by Lue and Starkman.

It has been shown that the systematic errors induced by the
multipolar expansion of the classical part of the terrestrial
gravitational potential rule out the possibility, suggested by
Ciufolini, of using the perigee of the proposed LARES/WEBER-SAT
satellite in order to test the  Dvali, Gabadadze and Porrati
gravity by measuring the Lue-Starkman pericentre precession.
Indeed, according to the EIGEN-CG01C Earth gravity model from the
dedicated CHAMP and GRACE missions, the errors induced by the
mismodelled even zonal harmonics of the multipolar expansion of
the Earth's gravitational potential  would be up to three orders
of magnitude larger than the LS precession of interest which
amounts to $\mp 4\times 10^{-3}$ mas yr$^{-1}$ only. Moreover,
also the observational accuracy, evaluated with optimistic
assumptions, would not be sufficient to measure such an effect
with the LARES perigee.

On the contrary, from the ranging data to Mars and Mercury it
might be possible, in a near future, to try to test the LS
precession, which amounts to $\mp 4\times 10^{-1}$ mas cy$^{-1}$,
with a suitable linear combination of the residuals of the
perihelia and the nodes of Mars and Mercury. Indeed, the
perihelion advance of Mars, which is the currently best tracked
planet of the Solar System, has recently been measured with a
formal standard deviation of $1.5\times 10^{-2}$ mas over a time
span of 92 years. The main systematic errors of gravitational
origin, which severely affects a single perihelion rate, are
instead cancelled out by the proposed linear combination. The
obtainable accuracy could roughly be evaluated to be of the order
of 41$\%$ at 1-sigma.

\section{Appendix: The impact of the asteroids on the Mars orbital
motion}\lb{asteroids} The bodies of the asteroid's belt induce
both secular and long-period perturbations on the orbital elements
of Mars whose amplitudes are proportional, among other things, to
their masses. Since they are relatively poor determined with
respect to those of the major planets of the Solar System the
mismodelled part of the orbital perturbations induced by them
might become relevant also because their nominal size is larger
than the LS effect. It is important to note that the periods of
some of the most powerful perturbations amount to tens and even
hundreds of years: it would be a major limiting factor for
analysis covering relatively short time spans of available data.

In the framework of the linear Lagrange-Laplace perturbation
theory the equation for the rate of the longitude of the
perihelion is\eqi \rp{d\varpi}{dt}=\rp{\sqrt{1-e^2}}{na^2
e}\derp{\mathcal{R}}e+\rp{\tan(i/2)}{na^2\sqrt{1-e^2}}\derp{\mathcal{R}}i,\eqf
where $\mathcal{R}$ is the disturbing function which accounts for
the effect of a third body, whose parameters are denoted with $'$,
on a given planet, whose parameters are denoted without $'$. It is
\eqi\mathcal{R}=GM^{'}\left(
\rp{1}{|\boldsymbol{r}-\boldsymbol{r^{'}}|}-\rp{\bf{r}\cdot\boldsymbol{r^{'}}}{{r^{'}}^3}
\right).\lb{dist}\eqf \Rfr{dist} can be expanded as a multiple
Fourier series with respect to the six orbital parameters of the
perturbed and perturbing bodies $\lambda, \varpi, \Omega,
\lambda^{'}, \varpi^{'}, \Omega^{'}$, where $\lambda $ is the mean
longitude \eqi
\mathcal{R}=\sum_{jj^{'}ll^{'}mm^{'}}F_{jj^{'}ll^{'}mm^{'}}\cos(j\lambda+j^{'}\lambda^{'}
+l\varpi+l^{'}\varpi^{'}+m\Omega+m^{'}\Omega^{'}).\lb{cazonga}\eqf
In turn, the coefficients $F_{jj^{'}ll^{'}mm^{'}}$, which depend
on $a,\ a^{'},\ e,\ e^{'},\ i,\ i^{'}$, can be expanded as a
Taylor series with respect to $e,\ e^{'},\ \sin i,\ \sin i^{'}$
due to the smallness of the eccentricities and the inclinations of
the Solar System bodies \eqi
F_{jj^{'}ll^{'}mm^{'}}=\sum_{r_1=0}^{\infty}
\sum_{r_2=0}^{\infty}\sum_{r_3=0}^{\infty}\sum_{r_4=0}^{\infty}C_{r_1
r_2 r_3 r_4 }e^{|l|+2r_1}{e^{'}}^{|l^{'}|+2r_2}\sin^{|m|+2r_3} i
\sin^{|m^{'}|+2r_4} i^{'}.\eqf In it the coefficients $C_{r_1 r_2
r_3 r_4 }$ depend on $a$ and $a^{'}$. In \rfr{cazonga} the so
called short-period term are those which contain the mean
longitudes $\lambda$ and $\lambda^{'}$, while the secular terms
are those which do not contain them, i.e. for which $j=j^{'}=0$.
In regard to the latter ones, it turns out that
$\left\langle{\mathcal{R}}\right\rangle$, averaged over the mean
longitudes, only contain the terms 1, $e^2$, ${e^{'}}^2$, $\sin^2
i$, $\sin^2 i^{'}$, $ee^{'}\cos(\varpi-\varpi^{'})$, $\sin i\sin
i^{'}\cos(\Omega-\Omega^{'})$.

Let us now specialize these general concepts to the Mars-asteroids
system.
\begin{itemize}
  \item According to \ct{jim}, the nominal amplitude of the secular perturbation on the longitude of the
  perihelion
  of Mars induced by the major asteroids amount to $3\times 10^{-2}$
  \asec. The three most massive bodies (1) Ceres, (2) Pallas and (4) Vesta are responsible of roughly 80$\%$
  of such perturbation, i.e. $\sim 2.4\times 10^{-2}$ \asec.
  The latest reported measurements of the masses
  of the asteroids \ct{pitast} are precise up to $\sim
  0.07-0.3\%$, so that the mismodelled part of the Mars perihelion
  secular precession due to the three major asteroids amounts to
  $\sim 1\times 10^{-4}$ \asec. The masses of the smaller asteroids like (3) Juno, (7)
  Iris and (324) Bamberga are known to a $\sim 1\%$ level, so that
  the mismodelled part of their perturbation may be evaluated as
  $\sim 2\times 10^{-4}$ \asec.
  The impact of the asteroid ring, i.e. the ensemble of the minor
asteroids which can be modelled as due to a solid ring in the
ecliptic plane \ct{kras}, can be worked out, e.g., with the
Lagrangian approach and the disturbing function of the Appendix of
\ct{kras}. By using the values of \ct{pitast} for the ring's
radius and mass it turns out that the secular perturbation on
$\dot\varpi_{\rm Mars}$ amounts to $2.8\cdot 10^{-3}$ \asec, with
an uncertainty of $3\cdot 10^{-4}$ \asec.
\item In regard to the short-period perturbations, according to \ct{jim}, the terms of \rfr{cazonga}
   containing also the mean longitudes are, in principle, not
   negligible because of the occurrence of many orbital resonances
   which induce harmonic perturbations with periods tens or even
   hundreds of years long. The focus in \ct{jim} is on the perturbations of the mean
   longitudes: the nominal values of their amplitudes are of the
   order of $10^{-3}-10^{-4}$ \asec (see Table IV of \ct{jim}). The current
   accuracy in our knowledge of the asteroids masses makes the
   mismodelled part of such perturbations smaller than the
   investigated braneworld effect: moreover, it should also be pointed out
   that the perturbations in the mean longitudes, which depend on the square of their
   periods, are larger than those in the other orbital elements,
   which, instead, depend linearly on their periods. These
   considerations would suggest that the resonant harmonic
   perturbations on the Martian perihelion due to the asteroids
   should not compromise the recovery of the LS effect. However, a
   careful and more detailed analysis of such an important topic
   is needed.
\end{itemize}

Concerning the error in the observational determination of the
Martian perihelion rate, according \ct{jim} in the DE 118 and DE
200 JPL ephemerides it was $6\times 10^{-3} $ \asec; with the
recent EPM2004 IAA (RAS) ephemerides it amounts to $1\times
10^{-4}$ \asec (E.V. Pitjeva, personal communication 2004).

\section*{Acknowledgements}
I am grateful to E. V. Pitjeva  and E. M. Standish  for their help
and clarifications. Thanks also to G. Starkman for his useful
remarks, to T.M. Eubanks for the important material sent to me and
for relevant discussions on the impact of the asteroids on the
Martian orbit and to J. Williams for his clarifications on the
possibilities of LLR.

\end{document}